\def\sl{\slshape}

\def\VE{\vfill\eject}
   
\input BoxedEPS
\SetTexturesEPSFSpecial
\HideDisplacementBoxes

\newcommand{\ci}{\cite}
\newcommand{\lab}{\label}
\newcommand{\eq}{\eqref}
\def\bl{\renewcommand{\baselinestretch}} 
\def\cl{\centerline}
\def\bib#1{\bibitem[#1]{#1}}

\def\la{\langle\, }

\def\ra{\,\rangle}

\def\0#1{{(#1)}}
\def\1#1{{\hat #1}}
\def\2#1{{\tilde #1}}
\def\3#1{{\boldsymbol#1}}
\def\4#1{{\mathbb#1}}
\def\5#1{{\cal#1}}
\def\6#1{_{\scriptscriptstyle#1}}
\def\7#1{{\bar#1}}
\def\8{\infty}
\def\9#1{^{\scriptscriptstyle#1}}
\def\/#1{{\bf#1}}
\def\;#1{{\breve#1}}

\def\bh#1{{\boldsymbol{\hat{#1}}}}

\def\v{\hskip.5pt \9\sharp} 
\def\v{\hskip.5pt\9\sharp} 
\def\w{\hskip.5pt\9\flat}

\def\d{\delta} 
\def\e{\varepsilon} 

\def\f{\phi} 
\def\vf{\varphi} 
\def\g{\gamma}
 
\def\i{\iota}

\def\m{\mu} 

\def\o{\omega} 
\def\p{\pi} 

\def\q{\theta} 
 
\def\r{\rho}
\def\vr{\varrho} 

\def\s{{\sigma}}

\def\D{\Delta}

\def\hb#1{{\qq\text{#1}\qq}}
\def\bull{$\bullet\ $}
\def\db{{d\kern-.8ex {^-}}}

\def\={\equiv} 
 
\def\cc#1{{{\mathbb C\hskip.5pt}^{#1}}}

\def\curl{\nabla\times} 

\def\div{\nabla\cdot }

\def\grad{\nabla} 

\def\im{{\,\rm Im}\ }  
\def\imp{\ \Rightarrow\ }
\def\i1#1{\int_{-\infty}^\infty d#1\, } 
\def\inv{^{-1}}

\def\pl{\partial}
\def\qq{\quad} 
\def\qqq{\qquad} 

\def\re{{\,\rm Re}\  }   
\def\rr#1{{{\mathbb R}^{#1}}}

\def\sr{\sqrt}

\def\sv#1{\vskip#1ex}

\def\frame#1#2{
    \cl{\vbox{\hrule height .3pt
    \hbox{\vrule width .3pt\kern 5pt
    \vbox{\kern 5pt
    \vbox{\hsize #1cm\noindent#2}
    \kern 5pt}
    \kern 5pt\vrule width .3pt}
    \hrule height 0pt depth.3pt}}}

\documentclass[12pt]{article}   
\usepackage{amssymb,amsmath}    
\begin{document}   

\hfill gr-qc/0108041
\sv1

\cl{Invited paper, \sl Workshop  on Canonical and
Quantum Gravity III}

\cl{Polish Academy of Sciences,
Warsaw, June 7-19, 2001. }

\cl{http://www.cft.edu.pl/GravityShop/index.html}

\sv3
    
\cl{\bf \large Distributional Sources for Newman's Holomorphic
Field}    

\sv3

{\cl{\bf  \large Gerald Kaiser\footnote{Supported by AFOSR 
    Grant  \#F49620-01-1-0271. \hfil\today}}
	 \cl{The Virginia Center for Signals and Waves}
   \cl{kaiser@wavelets.com $\bullet$\  www.wavelets.com}}

\sv4

\cl{\large\bf Abstract}
\sv3

In \ci{N73}, Newman considered the holomorphic extension
$\2{\3E}(\3z)$ of the Coulomb field $\3E(\3x)$ in $\rr3$. By
analyzing its multipole expansion, he showed that the real and
imaginary parts
\begin{align*}
\3E(\3x+i\3y)\=\re \2{\3E}(\3x+i\3y), \qq 
\3H(\3x+i\3y)\=\im \2{\3E}(\3x+i\3y),
\end{align*}
viewed as functions of $\3x$, are the electric and magnetic fields
generated by  a \sl spinning ring of charge \rm $\5R$. This represents
the EM part of the Kerr-Newman solution to the
Einstein-Maxwell equations \ci{NJ65, N65}. As already pointed out in
\ci{NJ65}, this interpretation is somewhat problematic since the
fields are double-valued. To make them single-valued, a branch cut
must be introduced so that $\5R$ is replaced by a \sl charged  disk
\rm $\5D$ having $\5R$ as its boundary. In the context of curved
spacetime, $\5D$ becomes a \sl spinning disk of charge and
mass \rm representing the singularity of the Kerr-Newman solution.

Here we confirm the above interpretation of $\3E$ and
$\3H$ without resorting to asymptotic expansions, by computing
the charge- and current densities  \sl directly \rm as distributions in
$\rr3$ supported in $\5D$. This shows  that $\5D$
\sl spins rigidly at the critical rate so that its rim $\5R$
moves at the speed of light. \rm

It is a pleasure to thank Ted Newman, Andrzej Trautman and Iwo
Bialy\-nicki-Birula for many instructive discussions, particularly
in Warsaw and during a visit to Pittsburgh.

\VE 

\bf 1. Introduction \rm
\sv1
The holomorphic extension of the Euclidean distance $r(\3x)$ in 
$\rr n$ was studied in \ci{K00}. Here we recall the basics for
$n=3$. Let
\begin{gather*}
\3z=\3x+i\3y\in\cc3, \qq  |\3x|=r,\qq |\3y|=a.
\end{gather*}
Define the \sl complex distance \rm in $\cc3$ as
\begin{gather*}
\2r(\3z)=\sr{\3z\cdot\3z}
=\sqrt{r^2-a^2+2i\3x\cdot\3y}\=p+iq\,,
\end{gather*}
whose branch points, for given $\3y\in\rr3$, form a \sl ring \rm in
$\rr3$:
\begin{align*}
\5R(\3y)=\{\3x\in\rr3: \2r=0\}=\{\3x\in\rr3: r=a,\,
\3x\cdot\3y=0\}.
\end{align*}
To make $\2r$ single-valued, choose
 the branch with $p\ge 0$, so that $\2r(\3x)=+r$. It then
follows that
\begin{align*}
0\le p\le r,\qqq -a\le q\le a,
\end{align*}
and $q$ has a  discontinuity  across
the \sl branch disk \rm
\begin{align*}
\5D(\3y)=\{\3x\in\rr3: p=0\}
=\{\3x\in\rr3: \3x\cdot\3y=0,\  r\le a\}
\end{align*}
given by
\begin{align}\lab{jump}
\3x\to\5D\9\pm(\3y)\=\5D(\3y)\pm i0\3y
\imp \2r\to\pm i\sqrt{a^2-\vr^2},
\end{align}
where $\vr$ is the radial variable on $\5D$.

$\5D$ can be deformed continuously to \sl any \rm surface
with $\pl\5D=\5R$, so it is really a  \sl flexible membrane
spanning $\5R$.  \rm

\sv2

The holomorphic Coulomb potential of  a unit charge and its
field are defined by
\begin{align}\lab{holcoul}
\2\f(\3z)=\frac 1{4\p\2r}, \qq
\2{\3E}(\3z)=-\grad\2\f=\frac{\3z}{4\p\2r^3}\,.
\end{align}
Because $\2\f$ is discontinuous across $\5D$, we specify that
$\grad$ be the \sl distributional gradient with respect to $\3x$.

Intuitively,  think of $\2\f(\3x+i\3y)$ and $\2{\3E}(\3x+i\3y)$ as
the Coulomb potential and Coulomb field at $\3x$ of a unit charge
displaced from the origin to the \sl imaginary location \rm $-i\3y$. 

We now show that this view is physically viable and
leads to some interesting and unexpected consequences. 

As in \ci{N73, NW74, NW74a} (see also \ci[Chapter 9]{K94}), define
the real fields $\3E, \3H$ on $\cc3$ by
\begin{align*}
\2{\3E}(\3z)=\3E(\3z)+i\3H(\3z).
\end{align*}
Regarding these as electric and magnetic fields, we will compute their 
source distributions.   

The \sl inhomogeneous \rm Maxwell equations
state that the charge and current densities are
\begin{align}\label{inhom}
\r(\3z)=\div\2{\3E}(\3z)=-\D\2\f, \qq
\3J(\3z)=-i\curl\2{\3E}(\3z),
\end{align}
while the \sl homogeneous \rm Maxwell equations require that $\r$
and $\3J$ be \sl real. \rm  The divergence and curl are the
distributional operators.

As $\2\f$ and $\2{\3E}$ are holomorphic when
$\3x\notin\5D(\3y)$,  we may compute $\r$ and $\3J$ outside of
$\5D$ by ordinary differentiation.  Since $\2\f$ is harmonic and
$\2{\3E}$ is a gradient, this gives
\begin{align*}
\r(\3x+i\3y)=0,\   \3J(\3x+i\3y)=\30 
\hb{for all } \3x\notin \5D(\3y).
\end{align*}  

The main goal of this paper is to  compute $\r$ and $\3J$ as 
distributions in $\3x$ supported on $\5D(\3y)$. This will be
done in two stages:

\bull First compute the  \sl surface \rm charge- and current densities 
on the \sl interior \rm of $\5D$. These will be seen to diverge on the
rim $\5R$.

\bull Then compute the \sl volume \rm densities
$\r(\3x+i\3y)$ and $\3J(\3x+i\3y)$ as  distributions  in
$\3x\in\rr3$, for any fixed $\3y\in\rr3$.

Of course the results of the second computation include those of the
first, but the first stage adds some clarity and makes contact with the
usual method of  bounday conditions.

\sv2

\bf 2. Interior Surface Densities \rm 
\sv1
As $\3x\to \5D\9\pm(\3y)$, \eq{jump} states  that
$\2r\to\pm i\sqrt{a^2-\vr^2}$, thus
\begin{gather*}
\2{\3E}=\frac{\3x+i\3y}{4\p\2r^3}
\to \pm\, i\frac{\3x+i\3y}{4\p (a^2-\vr^2)^{3/2}}\,,
\end{gather*}
The fields on $\5D\9\pm$ are therefore
\begin{align*}
\3E\9\pm= \frac{\mp\,\3y}{4\p (a^2-\vr^2)^{3/2}}, 
\qq \3H\9\pm= \frac{\pm\,\3x}{4\p  (a^2-\vr^2)^{3/2}}\,,
\end{align*}
with $\3E\9\pm$  normal to $\5D$ and $\3H\9\pm$ 
parallel to $\5D$. Their jumps are
\begin{align*}
\d\3E&\=\3E\9+-\3E\9-=-\frac{\3y}{2\p (a^2-\vr^2)^{3/2}}\\
\d\3H&\=\3H\9+-\3H\9-=\frac{\3x}{2\p(a^2-\vr^2)^{3/2}}\,.
\end{align*}
By placing infinitesimal pillboxes and loops around interior points
of $\5D(\3y)$ as in  \ci[p.~17]{J99}, we draw the following
conclusions:

\bull  Since the tangential component of $\3E$ and the
normal component of $\3H$ are continuous across $\5D$
(they actually vanish on $\5D\9\pm$),  the \sl homogeneous \rm
Maxwell equations are satisfied everywhere except possibly on $\5R$
(where $\3E\9\pm, \3H\9\pm$ diverge):
\begin{align*}
\curl\3E(\3x+i\3y)=\30 ,\qq \div\3H(\3x+i\3y)=0\qq
\forall\3x\notin\5R.
\end{align*} 
(We will see that these equations also hold on $\5R$.)

\bull The surface charge- and current densities on  $\5D$ are
\begin{align}\label{surf1}
\s&=\,\bh y\cdot\d\3E=-\frac{ a}{2\p (a^2-\vr^2)^{3/2}}\,,
\qqq \bh y\=\3y/a\\ 
\3K&=\bh y\times\d\3H
=\frac{\bh y\times\3x}{2\p(a^2-\vr^2)^{3/2}} 
=\frac{\vr\,\3e_\vf}{2\p
(a^2-\vr^2)^{3/2}}\,,\label{surf2}
\end{align}
where  $\3e_\vf$ is the unit vector along the azimuthal coordinate
$\vf$.

\bull The \sl velocity \rm of the charge at $(\vr, z,\vf)$ is
\begin{align}\lab{vel}
\3v=\frac{c\3K}\s=-\frac{c\vr\,\3e_\vf}a\,,
\end{align}
where we have inserted the speed of light $c$.  

\bull The charge moves at a constant angular velocity
$\3\o=-\bh y c/a$.

\bull Displacing the point charge to $+i\3y$ would thus give
\begin{align}\lab{angvel}\boxed{
\ \3\o=\bh y\, c/a. \ 
}\end{align}

\bull \sl  This rigid rotational motion is the maximum allowed
by relativity,  since $|\3v|\to c$ on the rim $\5R$.  \rm

\bull This is related to the singularity on $\5R$, since \eq{vel}
implies that
\begin{align}\lab{v/c}
\s=-\frac{ 1}{2\p a^2(1-v^2/c^2)^{3/2}}\,,\qq
\3K=\frac{\vr\,\3e_\vf}{2\p a^3(1-v^2/c^2)^{3/2}}\,.
\end{align}

However, the above surface densities cannot be the entire story, as
can be seen from the following considerations.

\bull It is  difficult to see how the surface densities $\s, \3K$ are
related to the distributions for a point source at the origin
($\3y=\30$), which ought to be $\r(\3x)=\d(\3x),\, \3J(\3x)=\30$.

\bull  In particular, according to \eq{surf1} the total charge with
\sl any \rm choice of $\3y$ would not be $e=1$, as we have
assumed, but
\begin{align*}
e(\3y)=\int_{{\5D(\3y)}}\s(\3x+i\3y)\vr\,d\vr\,d\vf
=2\p\int_0^a \s(\vr)\,\vr\,d\vr=-\8.
\end{align*}

This shows that $\s$ and $\3K$ are insufficient, and we must also
compute  the \sl singular \rm parts of the densities supported on
$\5R$. 

\sv2

\bf 3. The Volume Charge Distribution \rm 
\sv1
We first compute the charge density $\r$ as a distribution on $\rr3$,
using the \sl regularization \rm method introduced in \ci{K00}.

Given $\e>0$, the set 
\begin{align*}
\5D_\e(\3y)\=\{\3x\in\rr3: p=\e\}
\end{align*}
is an ellipsoid enclosing $\5D$, given in cylindrical coordinates
by
\begin{align*}
\frac{\vr^2}{a^2+\e^2}+\frac{z^2}{\e^2}=1.
\end{align*}
Define the \sl truncated field \rm by
\begin{align}\label{Ceps}
\2{\3E}_\e(\3z)=\q(p-\e)\2{\3E}(\3z),
\end{align}
where $\q$ is the Heaviside step function. Since  $\r$ and $\3J$
vanish outside of $\5D$, we have
\begin{align*}
\q(p-\e)\div\2{\3E}\=0, \qq \q(p-\e)\curl\2{\3E}\=\30.
\end{align*}
Define the  \sl $\e$-equivalent charge- and current distributions \rm
by
\begin{align*}
\r_\e(\3z)&\=\div\2{\3E}_\e=\d(p-\e)\grad p\cdot \2{\3E}\\
\3J_\e(\3z)&\=-i\curl\2{\3E}_\e=-i\d(p-\e)\grad p\times \2{\3E}.
\end{align*}
These are the distributions on $\5D_\e(\3y)$ needed to give
exactly the same field $\2{\3E}$ \sl outide \rm $\5D_\e$ while
giving a vanishing field \sl inside \rm $\5D_\e\,$. 

Since $\5D_\e$ has no boundary,  we  expect 
$\r_\e$ and $\3J_\e$ to be representable by smooth
\sl surface \rm densities on $\5D_\e\,$.

From \ci{K00}, we have (with $\2r^*$ the complex conjugate of
$\2r$) 
\begin{align*}
\grad p\cdot \2{\3E}
&=\frac{p\3x+q\3y}{\2r^*\2r}\cdot\frac{\3x+i\3y}{4\p\2r^3}
=\frac{p(r^2+q^2)+iq(a^2+p^2)}{4\p\2r^*\2r^4}\\
&=\frac{(p+iq)(a^2+p^2)}{4\p\2r^*\2r^4}
=\frac{a^2+p^2}{4\p\2r^*\2r^4}\\
\grad p\times \2{\3E}&=\frac{p\3x+q\3y}{\2r^*\2r}\times
\frac{\3x+i\3y}{4\p\2r^3}
=\frac{ip-q}{4\p\2r^*\2r^4}\,\3x\times\3y
=i\frac{\3x\times\3y}{4\p\2r^*\2r^4}\,.
\end{align*}
This gives (with $\2r=\e+iq$)
\begin{align}\label{regsources}
\r_\e=\d(p-\e)\,\frac{a^2+\e^2}{4\p\2r^*\2r^3}, \qqq
\3J_\e=\d(p-\e)\,\frac{\3x\times\3y}{4\p \2r^*\2r^3}.
\end{align}
The $\e$-equivalent sources are \sl
complex. \rm Their imaginary parts are  \sl fictitious magnetic
sources \rm  introduced by making the normal component of $\3H$
and the tangential component of $\3E$ discontinuous across
the ellipsoid $\5D_\e\,$. 

The \sl true \rm sources are then defined as
\begin{align*}
\r(\3x+i\3y)\=\lim_{\e\to 0}\r_\e(\3x+i\3y), \qq
\3J(\3x+i\3y)\=\lim_{\e\to 0}\3J_\e(\3x+i\3y),
\end{align*}
where the limit is in the sense of distributions in $\3x$, for
given $\3y$.

It will be found that $\r$ and $\3J$ are \sl real \rm distributions,
in agreement with the earlier observation that the \sl homogeneous
\rm Maxwell equations are satisfied.

As shown in \ci{K00},  $(p,q,\vf)$ are \sl oblate spheroidal
coordinates \rm in $\rr3$. The level sets of $p$  are oblate spheroids
$\5D_p$ like $\5D_\e\,$, and those of $q$ are  hyperboloids
$\5H_q$ orthogonal to $\5D_p$.

A test function in $\rr3$ can be expressed as
\begin{align*}
f(\3x)=f\v(p,q,\vf), \qqq p\ge 0, \ -a\le q\le a,\ 0\le\vf\le 2\p.
\end{align*}
In terms of $(p,q,\vf)$, the volume measure in $\rr3$ is
\begin{align}\lab{vol}
d\3x=\frac1a\,\2r^*\2r \,dp\,dq\,d\vf,
\end{align}
so the distributional action of $\r_\e$ on $f$ is  
\begin{align*}
\la \r_\e\,, f\ra&\=\int_\rr3 d\3x\ \r_\e(\3x+i\3y)\,f(\3x)\\
&=\frac1a\int_0^\8 dp\int_{-a}^a dq\,\2r^*\2r \,\r\v_\e(p, q)
\int_0^{2\p} d\vf\  f\v(p,q,\vf)\\
&=\frac{2\p}a \int_0^\8\2r^*\2r dp\int_{-a}^a dq\, \r\v_\e(p,q)
\7f\v(p,q),
\end{align*}
where  $\7f\v(p,q)$ is the \sl mean \rm of $f\v(p,q,\vf)$ over $\vf$. 

Inserting \eq{regsources} now gives
\begin{align*}
\la \r_\e\,, f\ra
&=\frac{a^2+\e^2}{2a}\int_{-a}^a dq\,
\frac{\7f\v(\e,q)}{(\e+iq)^2}\,.
\end{align*}
As claimed above, $\r_\e$ is represented by a smooth (but
complex) surface density on $\5D_\e\,$.

To obtain a finite limit as $\e\to 0$,  subtract and
add the linear Taylor polynomial $\7f\v(\e,0)+q\pl_q\7f\v(\e,0)$ in
the numerator:
\begin{align*}
\la \r_\e\,, f\ra
&=\frac{a^2+\e^2}{2a}\!\!\int_{-a}^a \!\!dq\,
\frac{\7f\v(\e, q)-\7f\v(\e,0)-q\pl_q\7f\v(\e,0)}{(\e+iq)^2}\\
&\qqq\qq+\frac{a^2+\e^2}{2a}\!\!\int_{-a}^a \!\!dq\,
\frac{\7f\v(\e,0)+q\pl_q\7f\v(\e,0)}{(\e+iq)^2}.
\end{align*}
In Section 4 of \ci{K00}, we found that
\begin{align}\label{lambda}
\int_{-a}^a \frac{dq}{(\e+iq)^2}&=\frac{2a}{a^2+\e^2}\\
\int_{-a}^a \frac{iq\, dq}{(\e+iq)^2}
&=\p-2\tan\inv(\e/a)-\frac{2\e a}{a^2+\e^2}\,.
\end{align}
Inserting these into the above expression and taking the limit
$\e\to 0$ now gives the action of the \sl true \rm charge distribution
as
\begin{align*}
\la \r\,, f\ra&=-\frac a2\int_{-a}^a dq\,
\frac{\7f\v(0, q)-\7f\v(0,0)-q\pl_q\7f\v(0,0)}{q^2}\\
&\qqq\qq+\7f\v(0,0)-i\frac{\p a}2\pl_q\7f\v(0,0).
\end{align*}
Since the test function is continuously differentiable and $(0, q,\vf)$
and $(0, -q, \vf)$ represent the same point on $\5D$, we have 
\begin{align*}
\7f\v(0,-q)=\7f\v(0,q) \hb{and} \pl_q\7f\v(0,0)=0,
\end{align*}
and
\begin{align}\label{q}
\ \la \r\,, f\ra =\7f\v(0,0)-a\int_0^a dq\,
\frac{\7f\v(0, q)-\7f\v(0,0)}{q^2}.\  
\end{align}
The cylindrical coordinates $(\vr, z)$ are related to $(p,q)$ by
\begin{align}\lab{cyl}
z\=\3x\cdot\bh y=\frac{pq}a, \qq
\vr\=\sr{r^2-z^2}=\frac{\sr{a^2+p^2}\sr{a^2-q^2}}a.
\end{align}
Writing
\begin{gather*}
f\w(\vr,z,\vf)\=f\v(p,q,\vf),\qq
\7f\w(\vr,z)\=\7f\v(p,q),
\end{gather*}
we  obtain the action of $\r$ in cylindrical coordinates as
\begin{align}\label{rhoreg}\boxed{
\ \la \r\,, f\ra =\7f\w(a,0)-a\int_0^a \vr\,d\vr\,
\frac{\7f\w(\vr,0)-\7f\w(a,0)}{(a^2-\vr^2)^{3/2}}.\ 
}\end{align}
The first term is the mean of $f$ on  $\5R$, while the second term is 
a \sl regularization \rm of the surface density $\s$  in \eq{surf1},
taking into account the singularity on $\5R$.

\bull Note that $\r$ is \sl real, \rm so there
are \sl no magnetic charges: \rm  $\div\3H\=0$.

\bull The subtraction in the second term 
makes $\r$ a \sl regularized distribution \rm in the sense of
\ci{GS64},  of the same type as the Cauchy principal value integral.  It
is  \sl not \rm an ordinary function to which values can be assigned at
points.  

\bull However, if  $f(\3x)=0$ on $\5R$, then \eq{rhoreg} becomes
\begin{align*}
\la \r\,, f\ra =-\frac a{2\p}\int_0^a
\frac{\vr\,d\vr}{(a^2-\vr^2)^{3/2}}\int_0^{2\p}
f\w(\vr,0,\vf)\, d\vf\,,
\end{align*}
reproducing the surface charge density $\s$ in  \eq{surf1}.

\bull For a \sl constant \rm test function, the second term in
\eq{rhoreg} vanishes. This term therefore represents a \sl single layer
of zero net charge. \rm 

\bull Taking $f(\3x)\=1$ gives the correct total charge as promised:
\begin{align*}
e(\3y)=
\int_\rr3\r(\3x+i\3y)\, d\3x=\la\r, 1\ra=1 \qqq \forall\3y\in\rr3.
\end{align*}

\bull Intuitively, therefore, the negative-infinite charge of $\s$  is
balanced by a positive-infinite charge on the rim, just so as
to give the correct total charge $e=1$.

\bull Unlike the singular expression \eq{surf1} for 
$\s$, the distribution \eq{rhoreg}
reduces to the point source as $\3y\to\30$:
\begin{align*}
\la \r\,, f\ra\to \7f\w(0,0)=f(\30),
\hb{hence}\r(\3x+i\3y)\to \d(\3x).
\end{align*}

\sv2

\bf 4. The Volume Current Distribution \rm
\sv1
Writing
\begin{align}\lab{xcyl}
\3x=\3\vr+z\bh y=\vr\,\3e_\vr+z\bh y
\end{align}
gives
\begin{align}\label{Jeps}
\3J_\e=\d(p-\e)\,\frac{\3x\times\3y}{4\p \2r^*\2r^3}
=-\d(p-\e)\,\frac{a\vr\,\3e_\vf}{4\p \2r^*\2r^3}\,.
\end{align}
Let $\3f(\3x)$ be a vector-valued test function in $\rr3$. Then
\begin{align*}
\3J_\e\cdot\3f=-a\d(p-\e)\,\frac{h\v(p,q,\vf)}{4\p \2r^*\2r^3},
\end{align*}
where
\begin{align*}
h\v(p,q,\vf)\=\vr\, \,\3e_\vf\cdot\3f\v(p,q,\vf)
\=\vr f\v_\vf(p,q,\vf).
\end{align*}
The distributional action of $\3J_\e$ on $\3f$ is therefore
\begin{align*}
\la\3J_\e\,, \3f\ra&\=\int_\rr3 d\3x\,\3J_\e\cdot\3f
=-\int_0^\8dp\int_{-a}^a dq\, \2r^*\2r
\d(p-\e)\,\frac{\7h\v(p,q)}{4\p \2r^*\2r^3}\\
&=-\frac12\int_{-a}^a dq\,\frac{\7h\v(\e,q)}{(\e+iq)^2}\,.
\end{align*}
As claimed, $\3J_\e(\3z)$ can be represented by a smooth (but
complex) surface current density on $\5D_\e(\3y)$. To
consider  $\e\to 0$,  subtract and add a linear Taylor
polynomial in the numerator:

\begin{align*}
\la\3J_\e\,, \3f\ra&
=-\frac12\int_{-a}^a dq\,
\frac{\7h\v(\e,q)-\7h\v(\e,0)-q\pl_q\7h\v(\e,0)} {(\e+iq)^2}\\
&\qqq\qq-\frac12\int_{-a}^a dq\,
\frac{\7h\v(\e,0)+q\pl_q\7h\v(\e,0)} {(\e+iq)^2}.
\end{align*}

Again, the smoothness of $h$ on $\5D$ implies 
\begin{align*}
\7h\v(0,-q)=\7h\v(0,q)\hb{and} \pl_q\7h\v(0,0)=0.
\end{align*}
Taking  $\e\to 0$ gives the action of the \sl true \rm current
distribution:
\begin{align*}
\la\3J\,, \3f\ra &=\int_0^a dq\,\frac{\7h\v(0,q)-\7h\v(0,0)}{q^2}
+\frac1a\,\7h\v(0,0)\,.
\end{align*}
This can be expressed in cylindrical coordinates as
\begin{align}\label{Jreg}\boxed{
\ \la\3J\,, \3f\ra=\7f\w_\vf (a,0)+
\int_0^a \vr d\vr\,
\frac{\vr\7f\w_\vf(\vr, 0)-a\7f\w_\vf(a, 0)} 
{(a^2-\vr^2)^{3/2}}.\ 
}\end{align}
The first term is the mean of $f\w_\vf$ on  $\5R$, while the second
term is  a regularization of the  surface 
current density $\3K$  in \eq{surf2}, taking into account the
singularity on $\5R$.

Similar remarks apply to those made earlier on $\r$: 

\bull $\3J$ is \sl real, \rm so there
are no magnetic currents: $\curl\3E\=\30$.

\bull The subtraction in the second term 
makes $\r$ a regularized distribution in the sense of
\ci{GS64}. 

\bull  If  $\3f=\30$ on $\5R$,  \eq{Jreg} reduces to 
\begin{align*}
\la\3J\,, \3f\ra=
\frac1{2\p}\int_0^a \vr d\vr\int_0^{2\p}d\vf\,
\frac{\vr \,\3e_\vf\cdot\3f}{(a^2-\vr^2)^{3/2}}\,,
\end{align*}
reproducing the surface current density 
\eq{surf2}. 

\bull Unlike the singular expression \eq{surf2} for $\3K$,  \eq{Jreg} 
has the correct point-source limit 
$\3y\to\30$:
\begin{align*} 
\la\3J\,, \3f\ra\to \7f\w_\vf (0,0)=0,
\hb{hence} \3J(\3x+i\3y)\to\30.
\end{align*} 

\sv2

\bf 5. The Magnetic Moment \rm
\sv1

The circulating current $\3J(\3z)$ generates a magnetic moment 
\begin{align*}
\3\m =\frac12\int_\rr3\3x\times\3J(\3x+i\3y)\,d\3x.
\end{align*}
Using \eq{xcyl} and \eq{Jeps} gives
\begin{align*}
\3x\times\3J_\e
&=\frac{\d(p-\e)\,\3x\times(\3x\times\3y)}{4\p\2r^*\2r^3}
=\frac{\d(p-\e)\,(az\3x-r^2\3y)}{4\p\2r^*\2r^3}\\
&=\frac{\d(p-\e)\,(az\3\vr+z^2\3y-r^2\3y)}{4\p\2r^*\2r^3}
=\frac{\d(p-\e)\,(az\3\vr-\vr^2\3y)}{4\p\2r^*\2r^3}.
\end{align*}
By \eq{vol} and \eq{cyl}, we have
\begin{align*}
\3\m_\e&\=\frac12\int_\rr3\3x\times\3J_\e\,d\3x
=\frac1{2a}\int_{-a}^a \frac{dq}{4\p(\e+iq)^2}
\int_0^{2\p}d\vf \,(az\3\vr-\vr^2\3y)\\
&=-\frac{(a^2+\e^2)\bh y}{4a^2}\int_{-a}^a dq
\,\frac{a^2-q^2}{(\e+iq)^2}.
\end{align*}
Using \eq{lambda} and letting $\e\to0$ gives
\begin{align*}
\3\m&\=\lim_{\e\to 0}\3\m_\e=-a\bh y =-\3y.
\end{align*}
The magnetic moment of a charge $e$ displaced to $+i\3y$ (instead
of $-i\3y$) is therefore
\begin{align}\lab{magmom}\boxed{
\ \3\m\=\frac12\int_\rr3\3x\times\3J(\3x-i\3y)\,d\3x=ec\,\3y.\ 
}\end{align}

\sv2

\bf 6. A ``Newtonian'' Gravitomagnetic Field? \rm 
\sv1
In Einstein's theory, a rotating mass generates a \sl
gravitomagnetic field \rm analogous to the magnetic field
generated by a rotating charge \ci{TPM86, CW95}. 
This manifests itself by ``dragging'' inertial frames near the
body along the direction of motion. 

We now repeat the above analysis with the Coulomb
potential replaced by Newton's potential and attempt to
interpret the results in terms of a ``Newtonian'' 
gravitomagnetic field.

Although the gravitomagnetic field has no strict counterpart in
Newtonian theory, recall that the rim $\5R$ of $\5D$ moves at the
speed of light, so our situation is essentially relativistic.
In any case, it is interesting to follow the mathematics even if
its physical significance is as yet unclear.

Begin by replacing Coulomb's potential with Newton's:
\begin{align*}
\f_e(\3x)=\frac e{4\p r}\qq \to\qq 
\f_m(\3x)=-\frac m{4\p r}.
\end{align*}
The holomorphic Newtonian potential and force field are
\begin{align*}
\2\f_m(\3z)=-\frac m{4\p \2r},\qq
\2{\3F}(\3z)=-\grad\2\f_m(\3z)=-\frac{m\3z}{\2r^3}.
\end{align*}
We want to investigate the physical significance of the real fields
\begin{align*}
\3F(\3z)=\re\2{\3F}(\3z),\qq \3G(\3z)=\im\2{\3F}(\3z),
\end{align*}
which are discontinuous on $\5D(\3y)$ and singular on $\5R(\3y)$.

By our earlier argument, we have the gravitational counterpart of the
homogeneous Maxwell equations:
\begin{align*}
\curl \3F(\3x+i\3y)\=\30 , \qq \div \3G(\3x+i\3y)\=0 \qqq
\forall \3x\in\rr3.
\end{align*}
\bull The conservative field  $\3F(\3x+i\3y)$ will be interpreted as
the Newtonian force field due to a mass distribution on $\5D(\3y)$.

\bull The divergenceless field $\3G(\3x+i\3y)$ is, by definition, the 
 ``Newtonian'' gravitomagnetic field.

 We define the \sl mass
density
\rm and \sl mass current density \rm by
\begin{align*}
\r_m(\3z)\=-\div \3F(\3z), \qqq \3J_m(\3z)\=-\curl\3G(\3z),
\end{align*}
where the sign is chosen so that  $\r_m(\3x)=m\d(\3x)$ in the case
of a real point source.

The electromagnetic and gravitational cases are related  by
\begin{gather*}
\3F(\3z)=-(m/e)\3E(\3z),\qqq \3G(\3z)=-(m/e)\3H(\3z)\\
\r_m(\3z)=(m/e)\r_e(\3z), \qqq \3J_m(\3z)=(m/e)\3J_e(\3z).
\end{gather*}
These relations can be used to transcribe all our results to the
gravitational case. 

In particular, placing a point mass $m$ at $i\3y$ transforms it into a
rigidly rotating disk $\5D$ with angular velocity
$\3\o=c\,\bh y/a$.  By \eq{magmom}, its  \sl spin angular
momentum
\rm is
\begin{align*} 
\3l\=\int_\rr3\3x\times\3J_m(\3x-i\3y)\,d\3x=
\frac m e\int_\rr3\3x\times\3J_e(\3x-i\3y)\,d\3x
=\frac{2m}e\,\3\m.
\end{align*}
The ``Newtonian'' gyromagnetic ratio  is therefore $\g\6N=e/2m$,
the classical  value for a distribution of  charged matter
with uniform charge-to mass density ratio. 

This differs markedly from the gyromagnetic ratio of a
Kerr-Newman solution \ci{DKS69}, which has the Dirac value
$\g=e/m$.

\sv2

\bf  7. Why do Imaginary Translations Generate Spin? \rm 
\sv1
We have seen that the \sl formal \rm imaginary translation of a point
charge from the origin to $i\3y$ has  \sl two real, physical effects: \rm

\sv1

\bull The point singularity \bf opens \rm to sweep out
an oriented disk $\5D$.

\bull  $\5D$ inevitably comes with its \sl maximum allowed  spin,
\rm rotating rigidly at the angular velocity $\3\o=(c/a)\,\bh y$ such
that the rim $\5R$ moves at the speed of light.

\bull The original charge is distributed uniformly on 
$\5R=\pl \5D$, while additional charges $Q_\5R=\8$ and
$Q_\5D=-\8$ with zero sum are distributed over $\5R$ (uniformly)
and $\5D$ (with surface density $\s$).

\sv1

How is this magic to be understood? Already in the 1950s, Ivor
Robinson was advocating the use of complex spacetime
in general relativity and electrodynamics, together
with \sl self-dual  fields \rm of the type $\3E+i\3H$; see
the Introduction by Rindler and Trautman in \ci{RT87}; also
\ci{T62, S58}.    

This led  in 1961 to the discovery \ci{R01} of what Penrose later
called the  \sl  Robinson congruence, \rm
instrumental in the formulation of Twistor theory; see \ci{P67} and 
\sl On the Origins of Twistor Theory \rm by Roger Penrose in
\ci{RT87}.  

After Kerr's landmark paper on the gravitational field of a spinning
mass  \ci{K63}, Newman and his collaborators showed that the Kerr
metric and its charged counterpart (Kerr-Newman metric) can be
``derived'' from the Schwarzschild metric using a complex
spacetime coordinate transformation \ci{NJ65, N65}.

\sv2

Newman et al. put ``derive'' in quotes  because they were at
the time unable to explain \sl why \rm these transformations should
lead to another solution of the  Einstein equation.

Much work has been done since then to develop and clarify this idea 
\ci{NW74,ABS75, F76}. One of the most beautiful
explanations of the connection between complex translations
 in \sl flat \rm spacetime and the generation of spin is
given by Newman and Winicour in \ci{NW74a}, but it seems to me
that not all of the mystery has been explained regarding the success
of this method in \sl curved \rm spacetime.      

I have pursued complex spacetime in a very different direction, as a
\sl relativistic phase space \rm on which to build quantum physics,
unaware at the time of all the above work  \ci{K90, K94}.  

There are reasons to hope that the analysis of the disk
singularity given here may be extended to curved spacetime, leading 
to \sl distributional energy-momentum tensors \rm  as  sources for
the Kerr and Kerr-Newman fields.

It should also be very useful to extend the above analysis from
static to radiating fields, in particular to the \sl electromagnetic
pulsed-beam wavelets  \rm proposed  for applications to radar and
communications \ci{K00a}. 

\sv4

\small
\bl {.8}

\end{document}